\documentclass{iopconfser}

\usepackage{amssymb,amsmath,amsthm,amsbsy,epsfig,color,graphicx,times}
\usepackage[ansinew]{inputenc}
\usepackage[english]{babel}
\usepackage{color}
\usepackage{braket}
\usepackage{tabularx}
\usepackage{array}

\begin{document}

\title{Dark matter induced by neutrino mixing and flavor vacuum condensate
probed by neutrino capture on tritium}

\author{Antonio Capolupo$^{1}$, Simone Monda$^{1}$,  Gabriele Pisacane$^{1}$, Raoul Serao$^{1}$,  and Aniello Quaranta$^{2}$}

\affil{$^1$Dipartimento di Fisica ``E.R. Caianiello'' and INFN gruppo collegato di Salerno, Universit\`{a} degli Studi di Salerno, I-84081
Fisciano (SA), Italy}

\affil{$^2$School of Science and Technology, University of Camerino,
Via Madonna delle Carceri, Camerino, 62032, Italy}

\email{capolupo@sa.infn.it, smonda@unisa.it, gpisacane@unisa.it, aniello.quaranta@unicam.it, rserao@unisa.it}

\begin{abstract}

We show that experiments designed to capture low energy neutrinos, like PTOLEMY are sensitive to the specific neutrino model. In particular, they might show a signature of the condensed flavor vacuum, featured in the flavor Fock space model, and allow  to test the hypothesis according to which  a possible dark matter component is determined by neutrino mixing \cite{capol2023}.

\end{abstract}

\section{Introduction}

Experimental validation of neutrino oscillations \cite{Osc1} - \cite{Osc4} provided  evidence for neutrinos having mass and for systems beyond the standard model of particles \cite{BSM1}- \cite{BSM44}. The accepted  theory of neutrino oscillations is the one proposed by Pontecorvo \cite{Pontecorvo1,Pontecorvo2}. However, despite general agreement on the   mechanism of particle oscillations, several unresolved questions remain in neutrino physics.
These include the origin of neutrino masses, whether neutrinos are Dirac or Majorana particles  and the appropriate field theoretical model for neutrino mixing.

Moreover, neutrinos play a key role in cosmology and astrophysics, since they are involved in leptogenesis models that explain the baryon asymmetry of the universe, and sterile neutrinos may contribute to dark matter \cite{CosmoNeutrino1,CosmoNeutrino2}. Dark matter, first introduced to explain galaxy rotation curves \cite{CDM2}, is one of the biggest unsolved mysteries in modern cosmology \cite{CDM1,CDM3}. It is modeled as a pressure-less perfect fluid, making up approximately $85\% $
of the total matter in the universe. Its   composition is   unknown. Several theories have been proposed for dark matter, including primordial black holes \cite{PDM1,PDM2}, supersymmetric particles \cite{SSDM}, axions \cite{ADM1} - \cite{ADM3}, and sterile neutrinos \cite{CosmoNeutrino2,SNDM}.

Recently, a new connection between particle mixing and cosmology has been explored \cite{curved2} - \cite{curved3}, by generalizing the   quantum field theory (QFT) model of neutrino mixing and boson mixing, based on the flavor Hilbert space \cite{FF1} - \cite{FF31}, to the case of curved space time \cite{curved1},  \cite{curved11}.

In refs.\cite{curved2}, \cite{curved3} it has been shown that the condensate structure of the  flavor vacuum associated to the neutrino mixing in curved background   contributes to the energy-momentum tensor of matter in a way equivalent to that of a pressure-less perfect fluid (a similar behavior occurs in the Minkowski metric  \cite{FDM} - \cite{FDM4}), implying that it could serve as a dark matter component.
Testing this theory is a significant challenge with current technology.

 The   flavor vacuum condensation also leads to QFT modifications in neutrino oscillation formulas, in flat \cite{FF11}, \cite{FF12} and in curved space \cite{curved1}, but these effects are   insignificant except for non-relativistic neutrinos, where $E_{\nu} \simeq m_{\nu}$,  while neutrino oscillations are typically observed at much higher energies, $E_{\nu} \gg m_{\nu}$. This excludes oscillation experiments as a tool to probe the condensation mechanism.
   However, recent  experiments such as PTOLEMY \cite{PTOLEMY}, \cite{Long2014} and KATRIN   \cite{Kat1} have focused on capturing the cosmic neutrino background, which is composed of non-relativistic neutrinos. Here we report the results of the paper \cite{capol2023}, in which we show that  low-energy neutrino experiments can differentiate between various QFT models of neutrino mixing  and
   indirectly test the hypothesis that the flavor vacuum contributes to dark matter.
   Indeed, we examine the impact of QFT condensation on the neutrino detection process, specifically neutrino capture on tritium, and show that the capture rate is influenced by the neutrino model used, including traces of the flavor vacuum condensate.

     The paper is organized as follows: in Section II, we explain  the link between neutrino mixing and dark matter. In Section III, we introduce  the inverse beta decay amplitude and discuss  four possible neutrino states according to known mixing schemes. In Section IV, we calculate the capture cross sections and rates for each neutrino state and show observable effects of flavor vacuum condensation. The final section presents the conclusions.

\section{Cosmological effects from the QFT of neutrino mixing}

  In the "flavor Fock space" approach \cite{FF1} - \cite{FF2}, flavor fields $\nu_e$, $\nu_{\mu}$ and $\nu_{\tau}$, which are produced and detected   in the weak interaction vertices neutrinos, are treated as the fundamental entities, and the usual mixing relations between neutrino states are extended to relations between quantum fields:
\begin{equation}\label{FieldMixing}
 \nu_e (x) = \cos \Theta \ \nu_{1} (x) + \sin \Theta \ \nu_2 (x) \ ; \ \ \ \ \   \nu_{\mu} (x) = \cos \Theta \ \nu_{2} (x) - \sin \Theta \ \nu_1 (x) \ .
\end{equation}
Here we have considered two flavors with mixing angle $\Theta$ for simplicity.
%
%
%
%
The mixing transformations   Eq.\eqref{FieldMixing} can be written in terms of a  generator $G_{\Theta} (t)$ as \cite{FF11}
as
$
\nu_e(x) = G_{\Theta}^{-1} (t) \nu_{1} (x) G_{\Theta} (t) $ and $ \ \nu_{\mu}(x) = G_{\Theta}^{-1} (t) \nu_{2} (x) G_{\Theta} (t)
$, and the flavor annihilators can be defined as
\begin{equation}\nonumber
 a_{\pmb{k},\nu_e}^{s} (t) = G_{\Theta}^{-1} (t) a_{\pmb{k},1}^s G_{\Theta}(t)=\cos \Theta \ a_{\pmb{k},1}^s + \sin \Theta \left(U_{\pmb{k}}^* (t) a_{\pmb{k},2}^s + \epsilon^{s} V_{\pmb{k}}(t)b^{s \dagger}_{-\pmb{k},2}\right)
 \end{equation}
and similar for $a_{\pmb{k},\nu_{\mu}}^s, b_{\pmb{k},\nu_{e}}^s, b_{\pmb{k},\nu_{\mu}}^s$, with $s= \pm 1/2$ and $\epsilon^s = (-1)^{s-\frac{1}{2}} $. The Bogoliubov coefficients \cite{Ume} - \cite{Ume2}: $U_{\pmb{k}}(t) = (u_{\pmb{k},2}, u_{\pmb{k},1})_t, V_{\pmb{k}}(t) = (u_{\pmb{k},1}, v_{\pmb{k},2})_t$ are given by the scalar products of modes with positive $u$ and negative energy $v$ and satisfy the relation: $|U_{\pmb{k}} (t)|^2 + |V_{\pmb{k}} (t)|^2 = 1$.
%
%
Mass and flavor  representations are unitarily inequivalent each other  \cite{FF11} and   the flavor annihilators define a time-dependent     flavor vacuum state $\ket{0_F (t)}= \ket{0_e (t)} \otimes \ket{0_{\mu} (t)}$: $a^{s  }_{\pmb{k},\nu_{\sigma}} (t) \ket{0_F (t)} = 0$, $\sigma = e, \mu$ which is different from the mass vacuum $\ket{0_M}= \ket{0_1} \otimes \ket{0_2}$.

The flavor vacuum assumes the structure of a condensate of particle-antiparticle pairs with definite masses \cite{FF11}.
The condensate number density is given by computing  the number expectation values on the flavor vacuum
$
N^F_{\pmb{k},i}  = \bra{0_F (t)} a_{\pmb{k},i}^{\dagger} a_{\pmb{k},i} \ket{0_F(t)} = \sin^2 \Theta |V_{\pmb{k}}|^2 = \bra{0_F (t)} b_{\pmb{k},i}^{\dagger} b_{\pmb{k},i} \ket{0_F(t)} \ ,
$
 for any $i=1,2$.

A generalization of the flavor Fock space approach to   curved spacetime has been analized in Refs.\cite{curved1}. In this case,
a non-vanishing energy momentum tensor associated to the flavor vacuum can be   studied within the semiclassical approach, where it appears on the right hand side of the Einstein field equations as a source term.
In particular, the contribution to the pressure   $p$ is zero for neutrino mixing, since one has: $\bra{0_F (t)} T^{i}_{i} \ket{0_F (t)} = 0$. On the other hand, the flavor vacuum is associated with a non zero energy density  \cite{curved2}, \cite{curved3}.

For a spatially flat Friedmann-Robertson-Walker metric with De Sitter evolution, at late times, the energy density takes the form \cite{curved2}

\begin{eqnarray}\label{EtaEtaFunction4}
  \nonumber \rho (\tau)  = \bra{0_F (t)} T^{0}_{0} \ket{0_F (t)}   &\simeq& i \sin^2 \Theta \sum_{\lambda} \int d^3 p |\Xi_p(\tau_0)|^2 \left(i \frac{H_0^3 \tau^3}{2 \pi^3} \right) \sum_{j=1,2} m_j \tanh \left(\frac{\pi m_j}{H_0} \right) \\
 \nonumber &+& \frac{i}{2} \sin^2 \Theta \sum_{\lambda} \int d^3 p \left[\Xi^*_p (\tau_0) \Lambda_p(\tau_0) \left(\frac{-im_1 H_0^3 \tau^3}{2 \pi^3 \cosh \left( \frac{\pi m_1}{H_0} \right)} \right) - c.c. \right] \\
  &-& \frac{i}{2} \sin^2 \Theta \sum_{\lambda} \int d^3 p \left[\Xi^*_p (\tau_0) \Lambda^*_p(\tau_0) \left(\frac{-im_2 H_0^3 \tau^3}{2 \pi^3 \cosh \left( \frac{\pi m_2}{H_0} \right)} \right) - c.c. \right] \  .
\end{eqnarray}
where $\tau$ is the conformal time, $H_0$ is the constant Hubble factor and $\Lambda$, $\Xi$ are the curved space counterparts of the Bogoliubov coefficients $U, V$.
We point out that, the components of the energy momentum tensor
  satisfy the typical equation of state of dust and cold dark matter \cite{CDM1,CDM2,CDM3}
\begin{equation}\label{FlatEquationofState}
 p = w \rho = 0 \ \ \longrightarrow \ \ w = 0   \ \ \ (\rho \neq 0) \ .
\end{equation}
 %
 %
%
We also note that  the energy density vanishes in absence of mixing $\Theta = 0$ and for $ \Xi  = 0$.

Recently, we also considered spherically symmetric spacetimes, and shown that  the flavor vacuum contributes as a Yukawa correction to the Newtonian potential.  This corrected potential may account for the flat rotation curves of spiral galaxies, and then, neutrino mixing could contribute to the cold dark  matter in the galaxies \cite{curved3}.

Considering the cosmological and astrophysical significance of this possibility, we propose an experimental test of the underlying theory.
We suggest using neutrino capture on tritium to probe the flavor Fock space model. This process is highly sensitive to extremely low-energy neutrinos, and potentially sensitive to quantum field theoretical corrections, which are  relevant for non-relativistic neutrinos ($E \sim m_j$). The quantum field corrections and the condensation density depend on the Bogoliubov coefficient $V_{\pmb{k}}$, which becomes negligible at high momenta, where $V_{\pmb{k}} \rightarrow 0$ and $U_{\pmb{k}} \rightarrow 1$, but is significant for energies close to the neutrino masses. The PTOLEMY experiment \cite{PTOLEMY}  aims to detect the cosmic neutrino background (C$\nu$B) consisting of very non-relativistic neutrinos. In the following, we demonstrate that if the flavor Fock space model is correct, the quantum field corrections will affect the neutrino capture rate. Such a rate  will vary depending on the neutrino model considered. Therefore, experiments designed to detect the C$\nu$B can help differentiate between various neutrino models and   validate the flavor vacuum hypothesis as a dark matter component.

\section{Hamiltonian and neutrino states}

  The basic reaction for neutrino capture is the inverse beta decay $\nu + n \rightarrow e^{-} + p$. Since to the neutrino energies involved are very low, much below the $W$ and $Z$ boson masses, the weak interaction Hamiltonian for the inverse beta decay    can be safely described with the current-current interaction Hamiltonian \cite{Long2014}:
\begin{equation}\label{Interaction Hamiltonian}
 H_I = \frac{G_F}{\sqrt{2}} V_{ud} \left[\bar{p}(x) \gamma_{\mu} \left( f(0) -g(0)\gamma^5 \right)n(x) \right] \left[\bar{e} (x) \gamma^{\mu} \left(1-\gamma^5\right)\nu_e (x) \right] + h.c. \ .
\end{equation}
Here $G_F = 1.166 \times 10^{-5} \mathrm{GeV}^{-2}$ is the Fermi constant, $V_{ud}$ is the CKM matrix element, $n(x), p(x), e(x), \nu_e (x)$ are respectively the neutron, the proton, the electron and the neutrino field, and $f(q)$ and $g(q)$ are nuclear form factors \cite{PDG}.   The tree level amplitude $\mathcal{A}$ is   defined as
\begin{equation}\label{Interaction Amplitude}
 S_{FI} = I_{FI} + (2\pi)^4 i \delta^4 (P_F^{\mu} - P_I^{\mu}) \mathcal{A}_{FI} = I_{FI} - i \left( \int d^4 x H_I (x) \right)_{FI} ,
\end{equation}
with   $FI$ denoting the matrix elements between the final state $F$ and initial $I$, $S$ is the S-matrix, $I$ is the identity matrix and the Dirac delta imposes total $4$-momentum conservation.
%
The initial and final states for the inverse beta decay are
\begin{equation}\label{Interaction States}
 \ket{I} = \ket{0_e} \ket{0_p} \ket{n_{P_n, s_n}} \ket{\nu_{e ; P_{\nu}, s_{\nu}}(I)} \ , \ \ \ \ \ \ket{F} = \ket{e_{P_e, s_e}} \ket{p_{P_{p},s_p}} \ket{0_n} \ket{0_\nu (F)}
\end{equation}
where the $P_j$ are momentum indices, the $s_j$ spin indices for $j=e,n,p,\nu$ and the tensor product is understood.

The form of the initial neutrino state $\ket{\nu_{e ; P_{\nu}, s_{\nu}}(I)} $ depends critically the mixing model considered.  We can distinguish four major cases:

\begin{enumerate}
 \item \emph{Decoupled Pontecorvo states}

 The neutrinos treated in the Ref.\cite{Long2014} are originally produced as Pontecorvo flavor states $\ket{\nu_e} = \cos \Theta \ket{\nu_1} + \sin \Theta \ket{\nu_2}$. However, due to the differing propagation speeds of $\nu_1$ and $\nu_2$, they rapidly decouple and   completely decohere into mass eigenstates.  Then, the inverse beta decay takes the form, $\nu_j + n \rightarrow e^{-} + p$.  The mixing effect   is only in the interaction Hamiltonian \eqref{Interaction Hamiltonian}, where the mixing matrix determines the fractions of $\nu_1$ and $\nu_2$ that interact as electron neutrinos.

 \item \emph{Pontecorvo states}

 Neutrinos are   produced and interact as Pontecorvo flavor states. The mixing is at the level of fields $\nu_e (x) = \cos \Theta \ \nu_1 (x) + \sin \Theta \ \nu_2 (x)$ and   therefore there is no decoherence.  The amplitude \eqref{Interaction Amplitude} is expressed in terms of the states     $\ket{\nu_e} = \cos \Theta \ket{\nu_1} + \sin \Theta \ket{\nu_2}$. In this case, the spinorial nature of the neutrino states is neglected.

 \item \emph{Pontecorvo-Dirac states}

The mixing is at the level of fields,  and the spinorial nature of the neutrino states is taken into account. The annihilator is $a_e^{\dagger}= (u_1, \nu_e (x))^{\dagger} \sim \cos \Theta a^{\dagger}_1 + \sin \Theta \  U a^{\dagger}_2 + \sin \Theta \ V b_2$, with $(,)$  the Dirac inner product and $b_2$ the anti-$\nu_2$ destruction operator and   $U$ and $V$ given by the scalar products $U \sim (u_1 , u_2), V \sim (u_1, v_2)$.
The action of $a^{\dagger}_e$  on the neutrino vacuum  produces the state $\ket{\nu_e} = \cos \Theta \ \ket{\nu_1} + \sin \Theta \ U \ket{\nu_2}$. This implies a unique neutrino vacuum state, the mass vacuum $\ket{0_F(t)}  \equiv \ket{0_M}$ annihilated by $a_1, a_2, b_1, b_2$. This implies that $
N^F_{\pmb{k},i}  = \bra{0_{M}} a_{\pmb{k},i}^{\dagger} a_{\pmb{k},i} \ket{0_{M}} = 0 \ ,
$
and the corresponding vacuum energy is equal to zero.


 \item \emph{Flavor Fock space states}

 The spinorial nature of neutrinos is taken into account, but the physical vacuum is the condensed flavor vacuum  $\ket{0_{F}(t)}$, belonging to a unitary inequivalent representation with respect to the mass vacuum $\ket{0_{M}}$.   The neutrino flavor state is of the form $\ket{\nu_e} = \cos \Theta \ \ket{\nu_1} + \sin \Theta \ U \ket{\nu_2} + sin \Theta \ V \ket{\overline{\nu}_2}  $.  flavor vacuum $\ket{0_{\nu,f}(t)}$ depends explicitly on time $t$, and this is the only case which features a condensed vacuum.
\end{enumerate}

Therefore, only in this last case one has a dark matter contribution from the vacuum of the theory.
 The last case is the most general, as all the others can be obtained by taking the appropriate limits.



  We compute the amplitude of eq. \eqref{Interaction Amplitude} and the corresponding cross section in the approach the flavor Fock space and from it we  derive all the other mixing schemes by doing the appropriate limits \cite{capol2023}.

\section{Amplitude and cross section}

We are interested to the unnormalized amplitude

\begin{equation}\label{Amplitude1}
 S_{FI} (T) = - i \left( \int_{-\frac{T}{2}}^{\frac{T}{2}} d x^0 \int d^3 x H_I (x) \right)_{FI}
\end{equation}
%
 Using the   states of Eq.\eqref{Interaction States} we have
\begin{eqnarray}\label{Amplitude2}
 \nonumber S_{FI}(T) &=& - i \frac{G_F}{\sqrt{2}} \int_{-\frac{T}{2}}^{\frac{T}{2}} d x^0 \int d^3 x \ V_{ud} \ \eta_{\alpha \beta} \bra{p_{P_p,s_p}} \bar{p}(x) \ket{0_p} \left[\gamma^{\alpha} \left( f - g\gamma^5\right)\right] \bra{0_n} n(x) \ket{n_{P_n,s_n}} \\
 &\times& \bra{e_{P_e,s_e}} \bar{e}(x) \ket{0_e} \left[ \gamma^{\beta} \left(1-\gamma^5 \right) \right] \bra{0_{\nu,f}(t_F)} \nu_e (x) \ket{\nu_{e; P_{\nu}, s_{\nu}}(t_I)} \ .
\end{eqnarray}
The matrix elements of the neutron, the proton and the electron states are computed straightforwardly. Since the flavor vacuum is time dependent, the neutrino matrix elements involves two   time arguments $t_I, t_F$ for the asymptotic states.
In any case, due to the small timescales of the weak interaction, we can  consider the approximation $t_F \simeq t_I$  and assume $t_F = t_I = 0$.
By using the flavor neutrino field  and the flavor annihilators, the neutrino matrix element is given by
\begin{eqnarray}\label{ExplicitState}
 \nonumber && \bra{0_{\nu, f}} \nu_e (x) \ket{\nu_{e; P_{\nu}, s_{\nu} }} = \\
 \nonumber && \sum_{r_{\nu}} \int \frac{d^3 k_{\nu}}{\sqrt{2 \pi^3}} e^{i \pmb{k}_{\nu} \cdot \pmb{x}} \left[u^{r_{\nu}}_{\pmb{k}_{\nu};1} \bra{0_{\nu, f}} \left \lbrace a^{r_{\nu}}_{\pmb{k}_{\nu}; \nu_e} (x^0) , a^{s_{\nu} \dagger}_{\pmb{p}_{\nu}; \nu_e}  \right \rbrace \ket{0_{\nu,f}} + v^{r_{\nu}}_{-\pmb{k}_{\nu};1} \bra{0_{\nu, f}} \left \lbrace b^{r_{\nu} \dagger}_{-\pmb{k}_{\nu}; \nu_e} (x^0) , a^{s_{\nu} \dagger}_{\pmb{p}_{\nu}; \nu_e}  \right \rbrace \ket{0_{\nu,f}} \right] = \\
 \nonumber && \frac{e^{i \pmb{p}_{\nu} \cdot \pmb{x}}}{\sqrt{2 \pi^3}} \Bigg[u^{s_{\nu}}_{\pmb{p}_{\nu};1}  \left(\cos \Theta \cos \Psi e^{-i E_{\nu;1}x^0} + \sin \Theta \sin \Psi \left(|U_{\pmb{p}_{\nu}}|^2 e^{-i E_{\nu;2}x^0} + |V_{\pmb{p}_{\nu}}|^2 e^{i E_{\nu;2}x^0}\right) \right) \\
  && + v^{s_{\nu}}_{-\pmb{p}_{\nu};1} \epsilon^{s_{\nu}} \sin \Theta \sin \Psi |U_{\pmb{p}_{\nu}}| |V_{\pmb{p}_{\nu}}|\left( e^{-i E_{\nu;2}x^0} - e^{i E_{\nu;2}x^0}\right)\Bigg]
\end{eqnarray}
We use the notation $\Theta$ and $\Psi$ to differentiate between the mixing angles in the fields and in the states.

Inserting the neutrino matrix element in Eq.\eqref{Amplitude2} and taking the infinite time limit $T \rightarrow \infty$ we find
\begin{equation}\label{Amplitude3}
 S_{FI} = \delta^4(p_{n}^{\mu} + p_{\nu,1}^{\mu} - p_{p}^{\mu}-p_{e}^{\mu}) M_{\nu_1} + \delta^4(p_{n}^{\mu} + p_{\nu,2}^{\mu} - p_{p}^{\mu}-p_{e}^{\mu}) M_{\nu_2} + \delta^4(p_{n}^{\mu} + \bar{p}_{\nu,2}^{\mu} - p_{p}^{\mu}-p_{e}^{\mu}) M_{\bar{\nu}_2} \ ,
\end{equation}
where $p_{\nu,j}^{\mu} \equiv (E_{\nu;j}, \pmb{p}_{\nu})$, $\bar{p}_{\nu,j}^{\mu} \equiv (-E_{\nu;j}, \pmb{p}_{\nu})$ and
\begin{eqnarray*}
 && M_{\nu_1} = - \frac{G_F}{\sqrt{2}} V_{ud} \eta_{\alpha \beta} \cos \Theta \cos \Psi R_{1}^{\alpha \beta} \ ; \ \  M_{\nu_2} = - \frac{G_F}{\sqrt{2}} V_{ud} \eta_{\alpha \beta} \sin \Theta \sin \Psi R_{2}^{\alpha \beta} \ ;\\
 &&   M_{\bar{\nu}_2} = - \frac{G_F}{\sqrt{2}} V_{ud} \eta_{\alpha \beta} \sin \Theta \sin \Psi \epsilon^{s_{\nu}} L_{2}^{\alpha \beta} \\
 && R^{\alpha \beta}_{j} = \bar{u}^{s_p}_{\pmb{p}_p} \gamma^{\alpha} (f-g\gamma^5)u^{s_n}_{\pmb{p}_n} \bar{u}^{s_e}_{\pmb{p}_e} \gamma^{\beta} (1-\gamma^5) u^{s_{\nu}}_{\pmb{p}_{\nu},j} \, , \ \ \ \  \ \ \ \ \ \ \ \ \  L^{\alpha \beta}_{j} = \bar{u}^{s_p}_{\pmb{p}_p} \gamma^{\alpha} (f-g\gamma^5)u^{s_n}_{\pmb{p}_n} \bar{u}^{s_e}_{\pmb{p}_e} \gamma^{\beta} (1-\gamma^5) v^{s_{\nu}}_{-\pmb{p}_{\nu},j} \ .
\end{eqnarray*}
We consider the neutron rest frame and neglect the proton recoil \cite{Long2014}, then the $4$-momenta are $p_{n}^{\mu} \equiv (m_n, 0), p_{p}^{\mu} \simeq (m_p,0), p_e^{\mu} \equiv (E_e , \pmb{p}_e), p_{\nu,j}^{\mu} \equiv (E_{\nu,j}, \pmb{p}_{\nu})$. We also set $\pmb{p}_e \cdot \pmb{p}_{\nu} = p_e p_{\nu} \cos \theta$, sum over the neutron and proton spins and average over the electron spin, in order to have the unpolarized cross section. Then
\begin{eqnarray}\label{Amplitude4}
\nonumber|M_{\nu_1}|^2 &=& 8 G_F^2 |V_{ud}|^2 \cos^2 \Theta \cos^2 \Psi m_p m_n E_e E_{\nu,1} \left[(1-2s_{\nu}v_{\nu})(3g^2 + f^2) + (v_{\nu}-2s_{\nu})(f^2 -g^2)v_{e} \cos \theta \right] \\
\nonumber |M_{\nu_2}|^2 &=&  8 G_F^2 |V_{ud}|^2 \sin^2 \Theta \sin^2 \Psi |U_{\pmb{p}_{\nu}}|^2 m_p m_n E_e E_{\nu,2} \left[(1-2s_{\nu}v_{\nu})(3g^2 + f^2) + (v_{\nu}-2s_{\nu})(f^2 -g^2)v_{e} \cos \theta \right] \\\nonumber
|M_{\bar{\nu}_2}|^2 &=& 8 G_F^2 |V_{ud}|^2 \sin^2 \Theta \sin^2 \Psi |V_{\pmb{p}_{\nu}}|^2 m_p m_n E_e E_{\nu,2} \left[(1-2s_{\nu}v_{\nu})(3g^2 + f^2) + (v_{\nu}-2s_{\nu})(f^2 -g^2)v_{e} \cos \theta \right]\\
\end{eqnarray}
where $v_j$ are   the velocities $v_j = \frac{p_j}{E_j}$.
 The differential cross section \cite{Weinberg} is thus given by \cite{capol2023}

\begin{eqnarray}\label{CrossSection1}
\nonumber \frac{d \sigma}{d \cos \theta} &=& \frac{1}{32 \pi} \frac{1}{m_n^2}\frac{p_e}{p_{\nu}} \left[ |M_{\nu_1}|^2 (E_{\nu,1}) + |M_{\nu_2}|^2 ( |E_{\nu,2}|) + |M_{\bar{\nu}_2}|^2  (- |E_{\nu,2}|) \right] \\ \nonumber
&=& \frac{G_F^2}{4 \pi}|V_{ud}|^2 \frac{m_p E_e p_e}{m_n} \Bigg[\left( A_1(s_{\nu}) (f^2 + 3g^2) + B_1(s_{\nu}) (f^2-g^2) v_e \cos \theta  \right)\frac{\cos^2 \Theta \cos^2 \Psi}{v_{\nu,1}}  \\
\nonumber &+& \left( A_2(s_{\nu}) (f^2 + 3g^2) + B_2(s_{\nu}) (f^2-g^2) v_e \cos \theta  \right)\frac{\sin^2 \Theta \sin^2 \Psi |U_{\pmb{p}_{\nu}}|^2}{v_{\nu,2}} \\
 &-& \left( A_{\bar{2}}(s_{\nu}) (f^2 + 3g^2) + B_{\bar{2}}(s_{\nu}) (f^2-g^2) v_e \cos \theta  \right)\frac{\sin^2 \Theta \sin^2 \Psi |V_{\pmb{p}_{\nu}}|^2}{v_{\nu,2}} \Bigg] \ .
\end{eqnarray}

Here the functions $A$ and $B$ are defined as
 $
A_{j}(s_{\nu}) = 1 - 2 s_{\nu} v_{\nu,j}$,
$
 B_{j}(s_{\nu})  = v_{\nu,j} - 2s_{\nu}$,
  $
  A_{\bar{j}}(s_{\nu}) = 1 + 2 s_{\nu} v_{\nu,j}$
 and  $
  B_{\bar{j}}(s_{\nu}) = -v_{\nu,j} - 2s_{\nu}$ ,
   in terms of the (positive) neutrino velocities $v_{\nu,j} = \frac{p_{\nu}}{|E_{\nu,j}|}$. Notice that the integral over $\cos \theta$  cancels all the $B$ terms.
Moreover, for the nonrelativistic neutrinos: $v_{\nu,j} = \frac{p_{\nu,j}}{E_{\nu,j}} \ll 1$, thus $A_{j} (s_{\nu}) = A_{\bar{j}} (s_{\nu}) \rightarrow 1$ and we can neglect the difference $v_{\nu,1} - v_{\nu,2} = \frac{p_{\nu}(m_2 -m_1)}{m_1 m_2} \ll 1$.
The capture cross section is then obtained multiply Eq. \eqref{CrossSection1} by, $v_{\nu,1} \simeq v_{\nu,2}$, and the Fermi function, $F(Z,E_e) = \frac{2 \pi \eta}{1-e^{-2\pi \eta}}$, with $\eta = \alpha \frac{E_e}{p_e}$.  It is given by
\begin{equation}\label{CrossSection2}
 \sigma = \frac{G_F^2}{2 \pi} |V_{ud}|^2 F(Z,E) \frac{m_p E_e p_e}{m_n} (f^2 + 3g^2) \left[\cos^2 \Theta \cos^2 \Psi + \sin^2 \Theta \sin^2 \Psi \left(|U_{\pmb{p}_{\nu}}|^2 - |V_{\pmb{p}_{\nu}}|^2 \right) \right] \ .
\end{equation}

This is the most general form for the capture cross section.

\begin{figure}[h]
\begin{center}
 \includegraphics[width=0.5\linewidth]{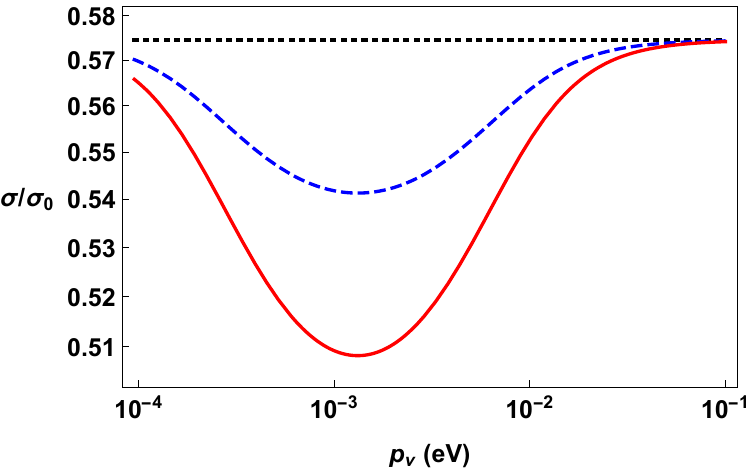}
 \caption{(color online): Plot of the ratio $\frac{\sigma (p_{\nu})}{\sigma_0}$ as a function of the neutrino momentum $p_{\nu}$ in the range $[10^{-4},10^{-1}] \  \mathrm{eV}$. The black dotted line corresponds to the Pontecorvo cross section (Eq. \eqref{CrossSection4}) $\sigma (p_{\nu}) = \sigma_P$; the blue dashed line is the Pontecorvo-Dirac cross section (Eq. \eqref{PDCrossSection} $\sigma (p_{\nu}) = \sigma_{PD}$, and the red solid line is the flavor Fock space cross section (Eq. \eqref{CrossSection5}) $\sigma (p_{\nu}) = \sigma_{F}$. The ratio of the cross section for decoupled neutrinos $\frac{\sigma_0}{\sigma_0}$ is $1$ by definition, and has not been plotted. The parameters have been chosen as $\sin^2 \Theta = \sin^2 \Theta_{12} = 0.307 $, $\Delta m^2 = \Delta m^2_{12} = 7.53 \times 10^{-5} \mathrm{eV}^2$ and $m_1 = 2 \times 10^{-4} \mathrm{eV}$.}
 \end{center}
 \end{figure}

From Eq. \eqref{CrossSection2}, we can give the capture cross section for the various mixing schemes considered:

\begin{enumerate}
 \item \emph{Decoupled Pontecorvo}

The cross section is   \cite{Long2014}

\begin{equation}\label{CrossSection3}
 \sigma_0 = \frac{G_F^2}{2 \pi} |V_{ud}|^2 F(Z,E) \frac{m_p E_e p_e}{m_n} (f^2 + 3g^2) \ .
 \end{equation}

 \item \emph{Pontecorvo}

 $U \rightarrow 1, V \rightarrow 0$ and $\Psi \rightarrow \Theta$, then

 \begin{equation}\label{CrossSection4}
 \sigma_P = \frac{G_F^2}{2 \pi} |V_{ud}|^2 F(Z,E) \frac{m_p E_e p_e}{m_n} (f^2 + 3g^2) \left[\cos^4 \Theta + \sin^4 \Theta \right] \ .
 \end{equation}

 \item \emph{Pontecorvo-Dirac}

 The cross section  depends on the neutrino momentum via $|U_{\pmb{p}_{\nu}}|$:

 \begin{equation}\label{PDCrossSection}
 \sigma_{PD} = \frac{G_F^2}{2 \pi} |V_{ud}|^2 F(Z,E) \frac{m_p E_e p_e}{m_n} (f^2 + 3g^2) \left[\cos^4 \Theta  + \sin^4 \Theta  |U_{\pmb{p}_{\nu}}|^2  \right] \ .
\end{equation}

   \item \emph{Flavor Fock space states}

We obtain

 \begin{equation}\label{CrossSection5}
  \sigma_F =  \frac{G_F^2}{2 \pi} |V_{ud}|^2 F(Z,E) \frac{m_p E_e p_e}{m_n} (f^2 + 3g^2) \left[\cos^4 \Theta  + \sin^4 \Theta  \left(|U_{\pmb{p}_{\nu}}|^2 - |V_{\pmb{p}_{\nu}}|^2 \right) \right] \ .
 \end{equation}


\end{enumerate}

The   cross sections of Eqs. \eqref{CrossSection3}, \eqref{CrossSection4}, \eqref{PDCrossSection} and \eqref{CrossSection5} represent the neutrino capture on a free neutron.   The capture cross sections on tritium $\nu_e + {^{3}H} \rightarrow {^{3}He} + e^{-}$ are found by replacing in the above equations   the nucleon masses $m_n , m_p$   with the masses of the atomic species $m_{{^{3}H}}$ and $m_{{^{3}He}}$, the form factors $f^2, 3g^2$   with the nuclear matrix elements $\langle f_F\rangle^2, (g_A/g_V)^2 \langle g_{GT}\rangle^2$, and modifying accordingly the electron kinetic energy $E_e - m_e$. Let us denote with
\begin{eqnarray*}
 \sigma_{n0} &=&  \frac{G_F^2}{2 \pi} |V_{ud}|^2 F(Z,E) \frac{m_p E_e p_e}{m_n} (f^2 + 3g^2) \\
 \sigma_{T0} &=&  \frac{G_F^2}{2 \pi} |V_{ud}|^2 F(Z,E) \frac{m_{{^{3}He}} E_e p_e}{m_{{^{3}H}}} (\langle f_F\rangle^2 + (g_A/g_V)^2 \langle g_{GT}\rangle^2)
\end{eqnarray*}
  the capture cross sections for decoupled neutrinos on free neutrons $\sigma_{n0}$ (Eq. \eqref{CrossSection3}) and on tritium $\sigma_{T0}$, respectively. Notice that the ratios between the cross section in a given scheme $\sigma =$ $\sigma_{P}$ (Pontecorvo), $\sigma_{PD}$ (Pontecorvo-Dirac), $\sigma_F$ (Flavor Fock space), and the reference cross section for decoupled neutrinos $\sigma_0$ are the same for the capture on free neutrons and on tritium, i.e. $\frac{\sigma_{n \alpha}}{\sigma_{n 0}} = \frac{\sigma_{T \alpha}}{\sigma_{T 0}}$ for each $\alpha =$ P, PD, F.
In Fig. 1 we plot such ratios     for neutrino momenta $p_{\nu} \in [10^{-4},10^{-1}] \  \mathrm{eV}$.

\begin{figure}[h]
\begin{center}
 \includegraphics[width=0.7\linewidth]{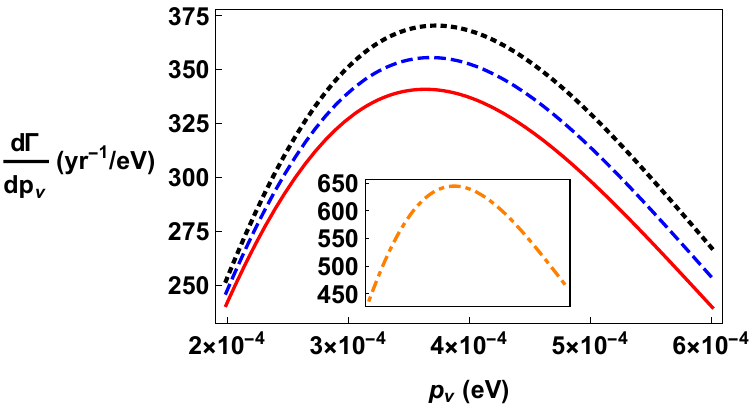}
 \caption{(color online): Plot of the differential capture rate (Eq. \eqref{capturerate}) $\frac{d \Gamma (p_{\nu})}{d p_{\nu}}$ as a function of the neutrino momentum $p_{\nu}$ in the range $[2\times 10^{-4},6 \times 10^{-4}] \  \mathrm{eV}$. The differential rate is of course adimensional, but is expressed in terms of $\mathrm{yr}^{-1}/\mathrm{eV}$ for ease of comparison with the previous results.
 The black dotted line corresponds to the Pontecorvo cross section (Eq. \eqref{CrossSection4}) $\sigma (p_{\nu}) = \sigma_P$, the blue dashed line is the Pontecorvo-Dirac cross section (Eq. \eqref{PDCrossSection} $\sigma (p_{\nu}) = \sigma_{PD}$ and the red solid line is the Flavor Fock space cross section (Eq. \eqref{CrossSection5}) $\sigma (p_{\nu}) = \sigma_{F}$. In the inset we plot the capture rate (orange dotdashed line) corresponding to the decoupled Pontecorvo cross section (Eq. \eqref{CrossSection3}). We have assumed a sample of tritium with total mass $M_S = 100 \  \mathrm{g}$. The other parameters have been chosen as in figure 1.}
 \end{center}
\end{figure}

The capture rate on a sample of tritium of mass $M_S$ is

\begin{equation}\label{TritiumCapturerate}
 d \Gamma = \sigma N_T d n_{\nu} \ ,
\end{equation}
where $N_T = \frac{M_S}{m_{{^{3}He}}}$ is the total number of tritium nuclei in the sample and $dn_{\nu}$ is the  differential  number density of neutrinos per degree of freedom.
In the sudden freeze-out approximation, the neutrino phase space distribution corresponds to the redshifted distribution function that existed at the decoupling epoch. At a redshift $z$, the number density is given by:
\begin{equation}\label{FDistribution}
 dn_{\nu} (z) = \frac{d^3 p(z)}{(2\pi)^3}\frac{1}{e^{\frac{p(z)}{T_{\nu (z)}}}+1} = \frac{p^2 (z) d p(z)}{2 \pi^2}\frac{1}{e^{\frac{p(z)}{T_{\nu (z)}}}+1}
\end{equation}
where   $p(z)= \frac{1+z}{1 + z_{FO}}p_{FO}$, $T_{\nu}(z)= \frac{1+z}{1 + z_{FO}}T_{\nu,FO}$ and $z_{FO} = 6 \times 10^{10}$, \cite{Long2014} and with $FO$ we denote the quantities at the freeze-out.
We   are analyzing the neutrino capture process which takes place in the present epoch. Therefore
we consider $z=0$ and we have
\begin{equation}
 dn_{\nu} (z=0) = dp_{\nu} \frac{p^2_{\nu}}{2 \pi^2} \frac{1}{e^{\frac{p_{\nu}}{T_{\nu}}}+1}
\end{equation}
where $p_{\nu} = p(z=0)$ and $T_{\nu} = T_{\nu} (z = 0) \simeq 0.168 \times 10^{-3} \mathrm{eV}$. The differential capture rate becomes
\begin{equation}\label{capturerate}
 \frac{d\Gamma}{dp_{\nu}} = N_T \frac{\sigma (p_{\nu}) p_{\nu}^2}{2 \pi ^2}\frac{1}{e^{\frac{p_{\nu}}{T_{\nu}}}+1}
\end{equation}
with the capture cross section $\sigma(p_{\nu})$.  In Fig.(2), we plot the differential capture rate for the various schemes discussed above.
%

From Fig.(2), we see that the capture rate for neutrinos is the lowest for the flavor Fock space states (red line in Fig.(2)). This   case   corresponds to the condensed flavor vacuum, and   to a possible dark matter component. In the plots, we assumed a sample of tritium with total mass $M_S = 100 g$ similar to the one used in PTOLEMY \cite{PTOLEMY}.

%
%
%

Notice that the behaviour of the capture cross sections is already contemplated in their two-flavor expressions, and only minimal differences arise from the introduction of a third flavor. Infact, in this case one has the cross section (compare with \eqref{CrossSection5}) \cite{capol2023}:
 %
%
\begin{equation}\label{CrossSection3flav}
  \sigma_{3F} =  \frac{G_F^2}{2 \pi} |V_{ud}|^2 F(Z,E) \frac{m_p E_e p_e}{m_n} (f^2 + 3g^2) \left[c_{12}^4 c_{13}^4  + c_{13}^4 s_{12}^4 \left(|U_{\pmb{p}_{\nu};12}|^2 - |V_{\pmb{p}_{\nu};12}|^2 \right)+ s_{13}^4 \left(|U_{\pmb{p}_{\nu};13}|^2 - |V_{\pmb{p}_{\nu};13}|^2 \right)\right] \ .
\end{equation}
The cross section for Pontecorvo-Dirac and   Pontecorvo formalisms are:
\begin{eqnarray}
 \nonumber  \sigma_{3PD} &=&  \frac{G_F^2}{2 \pi} |V_{ud}|^2 F(Z,E) \frac{m_p E_e p_e}{m_n} (f^2 + 3g^2) \left[c_{12}^4 c_{13}^4  + c_{13}^4 s_{12}^4 |U_{\pmb{p}_{\nu};12}|^2 + s_{13}^4 |U_{\pmb{p}_{\nu};13}|^2  \right] \\
 \sigma_{3P} &=&  \frac{G_F^2}{2 \pi} |V_{ud}|^2 F(Z,E) \frac{m_p E_e p_e}{m_n} (f^2 + 3g^2) \left[c_{12}^4 c_{13}^4  + c_{13}^4 s_{12}^4  + s_{13}^4   \right]
\end{eqnarray}
while the decoupled cross section is left unchanged $\sigma_{3,0} = \sigma_0$.

\section{Conclusions}

We have investigated the possibility for conducting experiments to test neutrino mixing and the influence of flavor vacuum on dark matter using low-energy neutrino capture experiments. We have explored various schemes of neutrino mixing and used the definition of flavor states in calculating the rate of neutrino absorption by tritium. Our findings indicate that the capture cross section and the capture rate are dependent on the definition of flavor states, particularly for very small neutrino momenta $p_{\nu} \simeq m_{\nu}$. This difference in values serves as a detectable indicator of the condensation of the flavor vacuum in the flavor Fock space model.

Experiments like PTOLEMY, designed to detect the cosmic neutrino background,   can potentially   differentiate among different mixing schemes and may provide an indirect evidence of the flavor condensate and dark matter contribution induced by neutrino mixing. A similar condensation mechanism is also predicted for mixed bosons  \cite{BosonMixing}, and investigations into the potential role of boson flavor vacuum as a component of dark energy have been conducted \cite{curved21}.
Future analysis on   $B$ meson decays $B^0 \rightarrow \phi \gamma$ and $B_s \rightarrow \rho^{0}(\omega)\gamma$ \cite{MesonMixing1},   \cite{MesonMixing2}, could potentially reveal observable signatures of boson flavor vacuum condensation \cite{BosonMixing}, \cite{FF31}, \cite{curved11} and of its contribution to the dark energy of the universe \cite{curved21}.

\section*{Acknowledgements}
Partial financial support from MIUR and INFN is acknowledged.
A.C. also acknowledges the COST Action CA1511 Cosmology
and Astrophysics Network for Theoretical Advances and Training Actions (CANTATA).

\end{document}